\input epsf.sty
\input mont.sty
\input dc_mont.sty

\hsize=16cm \vsize=21cm
\hfuzz=0.2cm
\tolerance=400
\parindent=18pt
\parskip=0pt plus 1pt
\def\ajnyp#1#2#3#4#5{
\frenchspacing{\smallsc #1}, {\sl#2}, {\bf #3} ({#4}) {#5}}
\smallskip
\rightline{hep-ph/99}
\rightline{\petit FTUAM 99-17, July, 1999}
\medskip
\noindent{{\twelverm NNLO Calculation of DIS;    
Precision Determination of the Strong Coupling
 Constant, Extraction of the Gluon Density, 
and Comments on ``Hidden" Gluinos}
\footnote*{\petit Presented 
to the Europhysics Conference QCD99, Montpellier, July 1999.}
}\vskip0.4cm
\noindent J. Santiago$^a$ and F. J. Yndur\'ain$^b$
\vskip0.3cm
\noindent {$^a$Departamento de F\'\i sica Te\'orica y del Cosmos,
 Universidad de  Granada, E-18071, Granada; \hb 
$^b$Departamento de F\'\i sica Te\'orica, C-XI,
Universidad Aut\'onoma de Madrid, 
Canto Blanco, 28049-Madrid}
\vskip0.6cm

{\petit \noindent
Calculations of  deep inelastic processes (DIS) to next-to-next to leading 
order are discussed. Fitting $ep$ experiment in the range 
$2.5\leq Q^2\leq 230\,\gev^2$, we find  the coupling 
 $\alpha_s(M_Z^2)=0.1163\pm0.0023$. We  also get  the gluon density  
 $xG(x,Q^2\simeq 10\,\gev^2)=0.51\, x^{-0.44}(1-x)^{8.1}$,  
and negative evidence for the existence of light gluinos}
\smallskip

\begindc{
\noindent{\fib 1. INTRODUCTION}
\smallskip
\noindent Deep inelastic scattering (DIS), in particular of electrons/muons 
on protons, constituted one 
of the first probes of hadron structure. The calculation of QCD-induced scaling 
violations in the structure functions\ref{1} yielded some of the 
earliest  checks of 
the quark-gluon theory of hadron interactions, as well as providing the first two loop  
determinations of the strong coupling constant\ref{2}. 

Let us set up some notation. Given the structure function  $F_2(x,Q^2)$ 
in $ep$ scattering, we split it into a singlet and a nonsinglet part, 
$$F_2(x,Q^2)=F_{NS}(x,Q^2)+F_S(x,Q^2).$$
For the second 
we have to consider also the gluon structure function, 
$F_G(x,Q^2)=xG(x,Q^2)$,  $G$  the gluon  density 
because  they mix. We  project the moments,
$$\mu_i(n,Q^2)=\int_0^1\dd x\,x^{n-2}F_i(n,Q^2),\quad i=NS,\,S,\,G$$
For  NS  the QCD NNLO evolution equation is
$$\eqalign{\mu_{NS}(n,Q^2)=\mu_{NS}(n,Q^2_0)
\left(\dfrac{\alpha_s(Q_0^2)}{\alpha_s(Q^2)}
\right)^{-\gamma_{NS}^{(0)}(n)/2\beta_0}\cr
\times\dfrac{1+B_n^{(1)}\alpha_s(Q^2)/4\pi+B_n^{(2)}(\alpha_s(Q^2)/4\pi)^2}{1
+B_n^{(1)}\alpha_s(Q^2_0)/4\pi+B_n^{(2)}
(\alpha_s(Q^2_0)/4\pi)^2}.\cr}
$$
The $B$ may be written in tems of anomalous dimensions, $\gamma^{(N)}(n)$ and 
Wilson coefficients, $C^{(N)}(n)$ of order $N$: 
to NNLO we require $N=0,\,1,\,2$. 
(For the singlet  equations, 
see ref.~3.)

To compare the QCD predictions with experiment we need thus to evaluate 
$\gamma^{(N)}(n),\;C^{(N)}(n)$
{\sl and}
 invert the moments equations. This  can be done with 
Altarelli--Parisi equations; but for this 
we would need the corresponding kernels, known only to NLO.\ref{4}
 It is also possible to 
invert the equations analytically if the analytic form of
 the $\gamma^{(N)}(n),\;C^{(N)}(n)$
 is known for all $n$ (ref.~5 to LO, and  ref.~2 to NLO). 
To NLO the calculations are 
complete\ref{2,4,6}; to  
NNLO we  have the calculations of ref.~7 
that 
provide us with the Wilson coefficients; but the 
$\gamma^{(N)}(n)$ are only known for a few values of $n$. Indeed,  
the calculations of Larin et al.\ref{8} give those  
 corresponding to NS  
scattering for $n=1$; and the 
singlet and nonsinglet $\gamma^{(N)}(n)$ in 
electroproduction for $n=2,\;4,\;6,\;8$. Therefore, 
before comparing with experiment 
some work has to be done. For the {\sl nonsinglet} case, see ref.~9;
 we next briefly describe the method 
followed by us in the singlet case, ref.~3 to where we send for more details. 
We also present here, for the first time, the ensuing 
determination of the gluon density, as well as a few comments on 
the (negative) implications of our analysis for the existence of light gluinos.
\smallskip
\noindent{\fib 2. BERNSTEIN AVERAGES}
\smallskip
\noindent For a given value of $Q^2$ only a limited 
number of experimental points, covering a partial range of values of $x$, 
are available, so one cannot simply use the moments equations.
 A method devised to deal with such a situation  is that 
of averages with the (modified) Bernstein
 polynomials (modified because, since only even moments are known, we 
have to consider polynomials in the variable $x^2$); for  
details on the method, see refs.~3,~10, and work 
quoted there. We define these  polynomials for 
$k\leq n$ as
$$p_{nk}(x)=
\dfrac{2\Gammav(n+\tfrac{3}{2})}
{\Gammav(k+\tfrac{1}{2})\Gammav(n-k+1)}x^{2k}(1-x^2)^{n-k}.$$
The $p_{nk}(x)$  are positive and
 have a single maximum located at 
$\bar{x}_{nk}\sim k/n$. They are concentrated around this point, with 
a spread, $\lap x_{nk}\sim 1/n$
 (accurate expressions for $\bar{x}_{nk},\;\lap
 x_{nk}$  
can be easily calculated, or looked for in ref.~3);
 and they are normalized to unity, 
$\int^1_0\dd x\,p_{nk}(x)=1$: so, the integral
$\int_0^1\dd x\,p_{nk}(x)\varphi(x)$ 
represents an average of the 
function $\varphi(x)$ in 
the region $\bar{x}_{nk}-\tfrac{1}{2}\lap x_{nk}\lsim
 x\lsim\bar{x}_{nk}+\tfrac{1}{2}\lap x_{nk}$.
 The 
values of the function $\varphi(x)$ outside this interval contribute  little
 to the integral, as $p_{nk}(x)$ 
decreases to zero  quickly there. Finally, 
 using the binomial expansion it follows that 
the averages with the $p_{nk}$ of a function can be obtained in terms of 
its {\sl even} moments:
$$\eqalign{\varphi_{nk}\equiv\int_0^1\dd x\,p_{nk}(x)\varphi(x)=
\cr=
\dfrac{2(n-k)!\Gammav(n+\tfrac{3}{2})}{\Gammav(k+\tfrac{1}{2})\Gammav(n-k+1)}
\sum_{l=0}^{n-k}\dfrac{(-1)^l}{l!(n-k-l)!}\,\mu^{\varphi}_{2k+2l},\cr
}$$
and $\mu^{\varphi}_{2l}=\int_0^1\dd x\,x^{2l}\varphi(x)$. 
 We will thus consider our 
experimental input to be given by averages
$$F^{\rm(exp)}_{nk}(Q^2)\equiv
\int_0^1\dd x\,p_{nk}(x)F_2^{\rm(exp)}(x,Q^2),
\equn{(2.1)}$$
 $F_2^{\rm(exp)}$  the experimental structure function.\ref{11} 
The experimental points (and the theoretical fit) are shown in the figure.
\smallskip
\setbox5=\vbox{\epsfxsize 7truecm\epsfbox{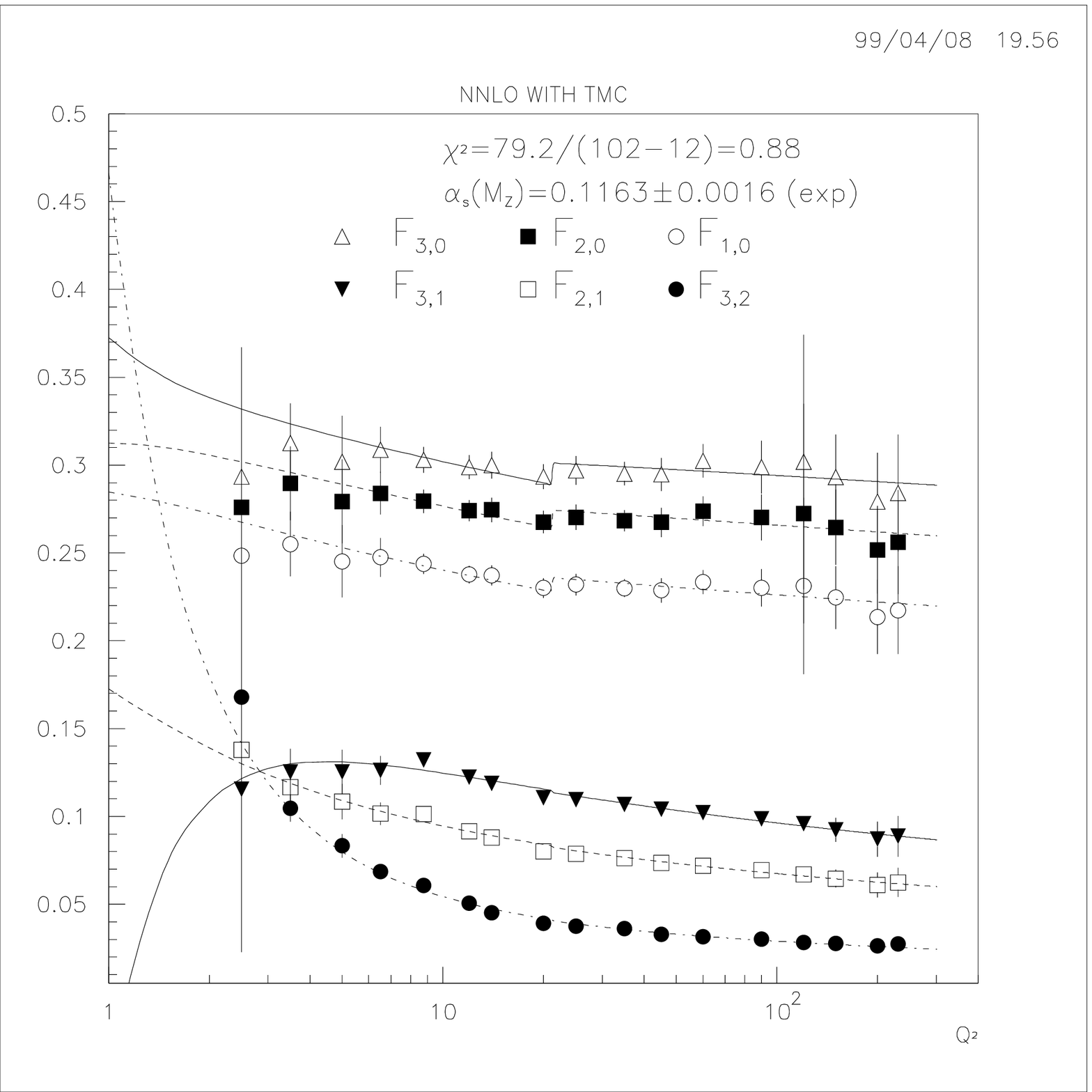}} 
\centerline{{\box5}}

\noindent{\fib 3. NUMEROLOGY}
\smallskip
\noindent We  present in Table 1 a compilation of the results obtained
 with our calculations at LO, NLO and 
NNLO, with TMCs 
(target mass corrections) taken into account; the fit to the data itself is shown,
 for the NNLO calculation (with TMC) in the previous  figure. Note 
the (small) jumps in the theoretical curves at the location of the mass 
threshold, $Q^2=m^2_b$; they   
 occur because 
we have joined the theoretical formulas from $n_f=4$ to $n_f=5$
 at that point, using the method of ref.~12, which is not exact.

\smallskip
\setbox0=\vbox{
\setbox1=\vbox{\petit\offinterlineskip\hrule
\halign{
&\vrule#&\strut\hfil#\hfil&\vrule#&\strut\hfil#\hfil&
\kern0.2em\vrule\kern0.2em#&\strut\kern0.2em#\kern0.2em&\kern0.2em\vrule#&\strut\quad#\cr
 height2mm&\omit&&\omit&&\omit&&\omit&\cr 
&\phantom{l}Order\phantom{l}&&\phantom{l}$\Lambdav(n_f=4)$&
&$\alpha_s(M_Z^2)$\kern.3em&&\chidof& \cr
 height1mm&\omit&&\omit&&\omit&&\omit&\cr
\noalign{\hrule} 
height1mm&\omit&&\omit&&\omit&&\omit&\cr
&LO&&\phantom{\big|}$ 215\pm73$&&$0.135\pm0.007$&&${212}/{102-12}\;$& \cr
\noalign{\hrule} 
height1mm&\omit&&\omit&&\omit&&\omit&\cr
&NLO&&\phantom{\big|}$ 282\pm40$&&$0.1175\pm0.0027$&&${80.0}/{102-12}\;$& \cr
\noalign{\hrule} 
height1mm&\omit&&\omit&&\omit&&\omit&\cr
&NNLO&&\phantom{\big|}$ 283\pm25$&&$0.1163\pm0.0016$&&${79.2}/{102-12}\;$& \cr
 height1mm&\omit&&\omit&&\omit&&\omit&\cr
\noalign{\hrule}}
\vskip.05cm}
\centerline{\box1}
\smallskip
\centerline{\petit Table 1}}
\box0 
We have 12 parameters: the twelve moments 
at the initial value, $\mu_i(n, Q^2_0)$, $n=2,\,4,\,6,\,8$ and 
$i=S, G, NS$, minus one moment, $\mu_G(2,Q^2_0)$ which is 
deduced from $\mu_S(2,Q^2_0)$ via the momentum sum rule; plus the QCD coupling 
parameter, $\Lambdav$.  
The initial value  is taken  $Q^2_0=8.75\,\gev^2$, and
 we evolve to all other values of $Q^2$, in the range
 $2.5\,\gev^2\leq Q^2\leq 230\,\gev^2$, with the 
QCD evolution equations; 
then construct the combinations entering the Bernstein averages, and fit 
the values of these obtained from experiment.  
In  Table 1, 
only  experimental ({\sl statistical}) errors of the fit 
are shown; systematic (theoretical) errors will 
be discussed below. 
The NLO corrections are  clearly seen in the 
fit: the 
\chidof\ decreases from a largish value of $\sim2.4$ to a very good $\sim0.89$.
 The fit is so good  at 
this order that there is  little room for improvement when going 
to NNLO; nevertheless, an improvement is seen. Not  
in the \chidof;  but   
 including NNLO corrections leads to 
 a noticeable gain both in the 
quality of the determination of the coupling, and in the stability of the fits.

Estimated systematic errors, originating  from various sources,
 are  shown for the NNLO calculation   
in Table 2.  
No TMC means that we have neglect  
target mass corrections. The 
corresponding  
error is {\sl not} included when evaluating the overall theoretical error,  
 since we {\sl do} take into account TMCs in 
our central value.  
 ``Interp." 
refers to the {\sl theoretical} errors inherent in the calculation of the 
integrals in (2.1) giving the experimental averages which arise 
because, for 
this calculation, it is necesary to interpolate the experimental points. 
We have used two 
different interpolation methods, one assuming a 
hard Pomeron (refs.~13) but 
with independent fits for every $Q^2$, which 
furnishes our central value, or 
using the MSRT98,\ref{14} that gives the estimated error.   
 HT means that we have 
taken into account  higher 
twists 
phenomenologically,   by adding, to
$\mu_{NS}(n,Q^2)$, the  
correction  
$\mu_{NS}^{HT}(n,Q^2)=n(a\Lambdav^2/Q^2)\mu_{NS}(n,Q^2).$
$a$ is free parameter whose  
 fitted value   is $a=-0.202\pm0.030$,
 a very reasonable number. 
 ``Quark mass effect" means that we 
cut off the $b$ quark threshold region, instead of matching thruough quark 
thresholds; 
the variation in Table~2 takes into account 
also the variations due to the uncertainty in the 
$m_b$ mass. $Q_0^2$ to $12\,\gev^2$ means 
that we take the input moments defined at this 
value of the momentum, $\mu_i(n,Q_0^2=12\,\gev^2)$, $i=S,\,G,\,NS$ 
instead of $Q_0^2=8.75\,\gev^2$ as was done to obtain the results of Table 1.
 Finally, {\sl N}NNLO is the estimated effect of the  
 (likely) 
larger sources of corrections of higher order in $\alpha_s$.

\setbox0=\vbox{
\medskip
\setbox1=\vbox{\petit\offinterlineskip\hrule
\halign{
&\vrule#&\strut\hfil#\hfil&\vrule#&\strut\hfil#\hfil&
\kern0.2em\vrule\kern0.2em#&\strut\kern0.2em#\kern0.2em&\kern0.2em\vrule#&\strut\kern0.2em#\cr
 height2mm&\omit&&\omit&&\omit&&\omit&\cr 
&\phantom{l}Source of error\phantom{l}&&\phantom{l}$\Lambdav(n_f=4,3\; {\rm loop})$&
&\phantom{l}$\lap\Lambdav$&&$\lap\alpha_s(M_Z^2)$\kern.3em& \cr
 height1mm&\omit&&\omit&&\omit&&\omit&\cr
\noalign{\hrule} 
height1mm&\omit&&\omit&&\omit&&\omit&\cr
&No TMC&&\phantom{\big|}$292$&&\hfil$9$\hfil&&$0.0006$& \cr
\noalign{\hrule} 
height1mm&\omit&&\omit&&\omit&&\omit&\cr
&\phantom{\big|}Interp.
\phantom{\big|}&&\phantom{\big|}$ 273$&&\hfil$10$\hfil&&$0.0007$& \cr
\noalign{\hrule} 
height1mm&\omit&&\omit&&\omit&&\omit&\cr
&HT&&\phantom{\big|}$292$&&\hfil$9$\hfil&&$0.0006$& \cr
\noalign{\hrule} 
height1mm&\omit&&\omit&&\omit&&\omit&\cr
&Quark mass effect&&299&&\hfil$16$\hfil&&$0.0010$& \cr
\noalign{\hrule}
height1mm&\omit&&\omit&&\omit&&\omit&\cr
&$Q_0^2$ to $12\;\gev^2$&&\phantom{\big|}$294$&&\hfil$11$\hfil&&$0.0007$& \cr
\noalign{\hrule}
height1mm&\omit&&\omit&&\omit&&\omit&\cr
&{\sl N}NNLO&&\phantom{\big|}$289$&&\hfil$6$\hfil&&$0.0004$& \cr
 height1mm&\omit&&\omit&&\omit&&\omit&\cr
\noalign{\hrule}}
\vskip.05cm}
\centerline{\box1}
\smallskip
\centerline{\petit Table 2}
}
\box0
\smallskip

Composing quadratically {\sl all}   
errors we find    
$$\eqalign{\Lambdav(n_f=4,\,\hbox{3 loop})=&283\pm25\;
(\hbox{stat.})\pm24\;(\hbox{syst.})\cr
=&283\pm35\;\mev;\cr
 \alpha_s^{(\rm 3\, loop)}(M_Z)
=&0.1163\pm0.0023.\cr} 
$$

In Table 3 we compare our results to previous determinations
\ref{15} for $\alpha_s(M_Z^2)$,
 to the NNLO level (but excluding $e^+e^-$ annihilations.) 
DIS means deep inelastic scattering, 
Bj stands for the Bjorken,  GLS for the Gross--Llewellyn~Smith 
sum rules. The $xF_3$ result is that of ref.~9.

The previously existing average,
 also taking into account NLO calculations, was
$\alpha_s(M_Z^2)=0.118\pm0.006;$
when including both our result and that of ref.~9 the new average
and error become
$$\alpha_s(M_Z^2)=0.1165\pm0.0016.$$

We discuss briefly how the calculation could improve. 
Adding the values of $\gamma(n)$ for a few more values of $n$ 
would allow us to extend the range and increase the precision of our evaluations: 
alredy two more moments would probably 
decrease the error in $\Lambdav$ in some 30\%. 
This is a difficult task, but it appears that Moch and Vermaseren are on 
the way to it (see the paper by S. O. Moch, these Proceedings; it 
is even possible that the anlytic expression be found for $\gamma(n)$).  

\setbox0=\vbox{
\medskip
\setbox1=\vbox{\petit \offinterlineskip\hrule
\halign{
&\vrule#&\strut\hfil#\hfil&\kern0.2em\vrule\kern0.2em#&
\strut\kern0.2em#\kern0.2em&\kern0.2em\vrule#&\strut\kern0.4em#\cr
 height2mm&\omit&&\omit&&\omit&\cr 
& Process&&${\textstyle\hbox{Average}\;
 Q^2\;{\rm or}}\atop{\textstyle Q^2\;\hbox{range}\;(\gev) }$&
& $\alpha_s(M_Z^2)$& \cr
 height1mm&\omit&&\omit&&\omit&\cr
\noalign{\hrule} 
height1mm&\omit&&\omit&&\omit&\cr
&\phantom{\Big|}  $\tau$ decays&&$1 - (1.777)^2$&&0.119&\cr
\noalign{\hrule}
&\phantom{\Big|}  $Z\to{\rm hadrons}$&&\hfil $(91.2)^2$\hfil&&0.124&\cr
\noalign{\hrule}
&\phantom{\Big|}  DIS; $\nu$, Bj&&\hfil 2.5\hfil&&$0.122^{+0.005}_{-0.009}$\phantom{l}& \cr
\noalign{\hrule}
&\phantom{\Big|}  DIS; $\nu$, GLS&&\hfil 3\hfil&&$0.115\pm0.006$\phantom{l}& \cr
\noalign{\hrule}
&\phantom{\Big|}  DIS; $\nu,\,xF_3$&&$5 - 100$&&$0.117\pm0.010$\phantom{l}& \cr
\noalign{\hrule}
&\phantom{\Big|}  DIS; $e/\mu p,F_2$&&$2.5 - 230$&&$0.1163\pm0.0023$\phantom{l}&\cr
\noalign{\hrule}
&\phantom{\Big|}  Average DIS&&$2.5 - 230$&&$0.1168^{+0.0019}_{-0.0020}$\phantom{l}&\cr
 height1mm&\omit&&\omit&&\omit&\cr
\noalign{\hrule}}
\vskip.05cm}
\centerline{\box1}
\smallskip
\centerline{\petit Table 3}
}
\box0

\noindent{\fib 4. THE GLUON STRUCTURE FUNCTION}
\smallskip
\noindent A spin-off from our results is that we also get 
moments of the gluon structure function, 
in particular at our starting value of $Q^2=Q^2_0$:
{\petit
$$\matrix{Q_0^2=8.75 &\quad Q_0^2=12\cr
\mu_G(2)=0.242041\pm0.003&0.2390\pm0.0015\cr
\mu_G(4)=0.0020\pm0.00023&0.0020\pm0.0055\cr
\mu_G(6)=0.00119\pm0.00050&0.00142\pm0.0013\cr
\mu_G(8)=(0.99\pm0.86)10^{-3}&(1.00\pm0.84)10^{-3}.\cr}$$
\medskip}

The results,  fairly stable,  
 do   {\sl not} fix 
the gluon density; to get it, extra assumptions have to be made. 
Following the hard Pomeron model  we take the functional form 
$$F_G(x,Q^2)\equiv xG(x,Q^2)=A_Gx^{-\lambda}(1-x)^\nu,$$
 fixing $\lambda=0.44$ (ref. 13) to avoid spureous minima.  
Fiting the moments gives
{\petit
$$\matrix{Q^2=8.75\;{\rm GeV}^2&Q^2=12\;{\rm GeV}^2&Q^2=8.75\&12\cr
A_G=0.52&A_G=0.51&A_G=0.51\cr
\nu=8.17&\nu=8.08&\nu=8.11\cr
\chi^2=5.8&\chi^2=2.4&\chi^2=9.0\cr}$$}
the last if making a global fit. The value of the 
 $\chi^2/{\rm d.o.f.}=9.0/(8-2)$
 in this last case  is quite reasonable: note  that 
$A_G$ and $\nu$ should depend on $Q^2$, so we are, in the last fit,  
obtaining an average. 

It is interesting to compare the value obtained for $A_G$ 
with the predicition of the hard Pomeron model. We write\ref{13}
$$\eqalign{F_S(x,Q^2)\simeqsub_{x\to0}\langle e_q^2\rangle
A_S(Q^2) x^{-\lambda},\quad \lambda=0.44,\cr
A_S(Q^2) =\langle e_q^2\rangle 
\alpha_s(Q^2)^{-d_+(1+\lambda)}B_S ,\;B_S=\hbox{const.},\cr}$$
$A_G/B_S=[d_+(1+\lambda)-D_{11}(1+\lambda)]/D_{12}(1+\lambda)
\simeq4.82$. Thus
$\langle e_q^2\rangle B_S\simeq2.1\times10^{-3}$, 
in the ballpark of the values found 
in ref.~13.
\smallskip
\noindent{\fib 5. COMMENTS ON ``HIDDEN" GLUINOS}
\smallskip
\noindent It is recurrently suggested  that  ``hidden" 
light gluinos could exist.\ref{16}
 They would alter the evolution 
of $\alpha_s$: in this respect, the agreement of our determinations with 
those made at smaller ($\tau$ decay) or 
 higher energy ($Z$ decay) 
provides strong evidence against their existence. Direct 
negative evidence is obtained as follows. Let  $\chidof(M)$ be 
the chi square per degree 
of freedom, taking into account {\sl only} experimental
 points with $Q^2\leq M^2$. 
If, at a given $M=M_0$ a channel for the production of 
particles not taken into account in the analysis 
opened, then   $\chidof(M)$ 
should jump at $M=M_0$, and would
 continue deteriorating for larger $M$, as we are fitting 
with  {\sl wrong} theoretical formulas:
  neglect of NLO corrections deteriorates the \chidof\ 
substantially, and gluinos  contribute already to LO.
 A plot  
 of $\chidof(M)$ at different $M$  
 is given in the figure (the full dots; the straight 
 line is the ideal value $\chidof(M)=1$). 
 The increase of the $\chidof(M)$ near 
$M=m_b$ is clearly seen, due to our 
approximate treatment of the threshold;
 but, since we are using  correct theoretical formulas 
 (that is, with $n_f=5$) above that value, $\chidof(M)$ 
decreases as we get far from $m_b$. 
This shows the effectiveness of the method.
\smallskip
\setbox5=\vbox{\epsfxsize 7truecm\epsfbox{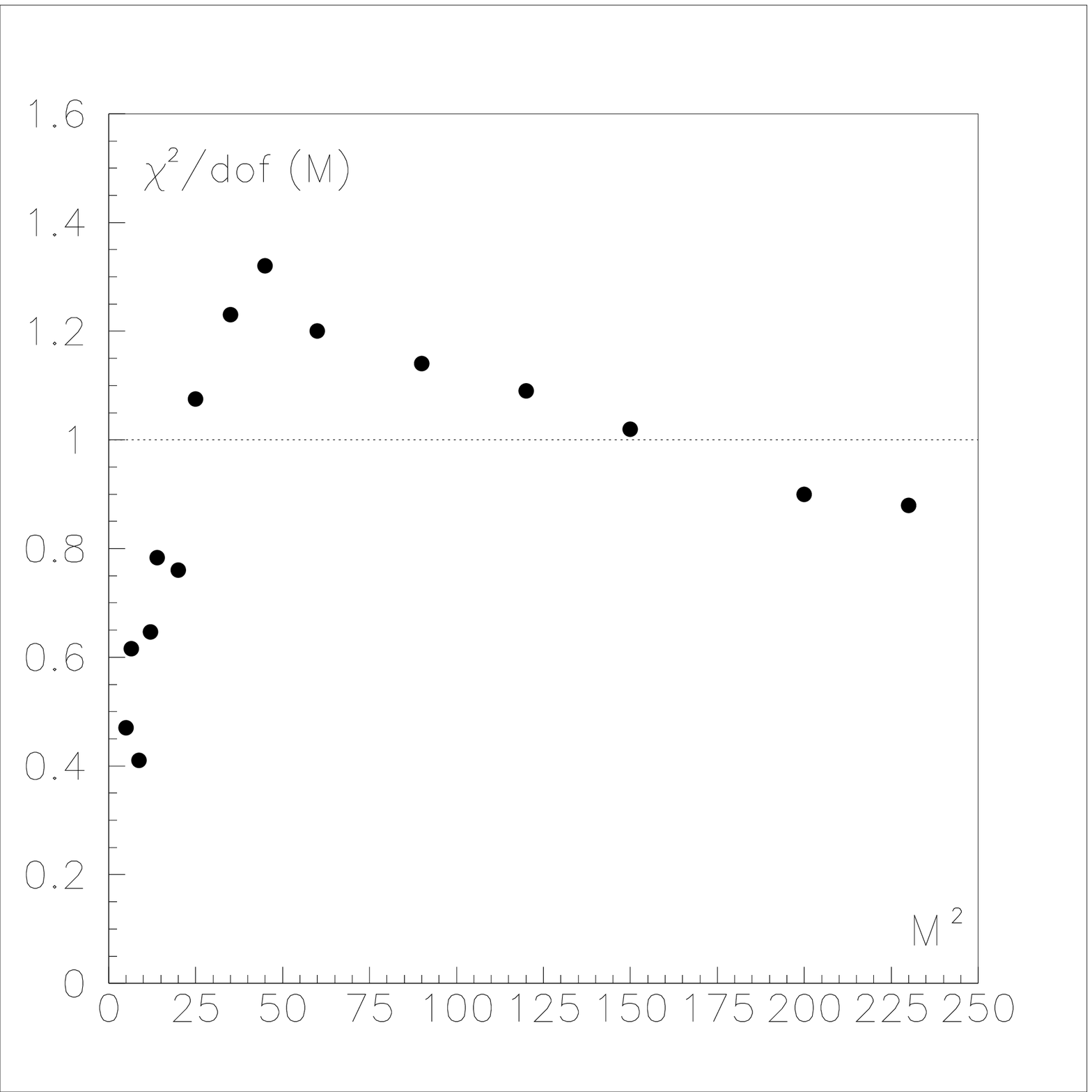}} 
\centerline{{\box5}}

No other step, or deterioration is seen in 
the whole range. Now, the presence of a particle with mass $M_0$ produces effects 
in  the Bernstein averages   
below $Q^2\simeq M_0^2$ because they involve integrals 
with the variable $x$. 
We are sensitive to energies up to   
$E_{\rm Max.}\simeq\sqrt{(1-\bar{x})Q^2_{\rm Max.}/\bar{x}}\simeq 30\,\gev$,
 and slightly above. So   
it follows that we can exclude gluinos with masses below $16\sim20\;\gev$.
 
\smallskip
\noindent{\fib REFERENCES}
\smallskip
{\petit
\item{1}{\ajnyp{D. J. Gross and F. Wilczek}{Phys. Rev.}{D9}{1974}{980}; 
\ajnyp{H. Georgi and H. D. Politzer}{Phys. Rev.}{D9}{1974}{416}; 
\ajnyp{I. Hinchliffe and 
C. H. Llewellyn Smith}{Nucl. Phys.}{B128}{1977}{93};
 \ajnyp{A.  De R\'ujula, H. Georgi
 and H. D. Politzer}{Ann. Phys. ({\rm NY})}{103}{1977}{315}.}
\item{2}{\ajnyp{A. Gonz\'alez-Arroyo,  C. L\'opez and 
 F.~J.~Yndur\'ain}{Nucl.
 Phys.}{B153}{1979}{161}; 
{\bf B159} (1979) 512; {\bf B174},
 (1980) 474.}
\item{3}{{\smallsc J. Santiago and F.~J.~Yndur\'ain}, {FTUAM 99-8/UG-FT-97/99}
 (hep-ph/9904344).}
\item{4}{\ajnyp{G. Curci, W. Furmanski
 and R. Petronzio}{Nucl. Phys.}{B175}{1980}{27}; 
 \ajnyp{W. Furmanski and R. Petronzio}{Phys. Lett.}{97B}{1980}{437}.}
\item{5}{\ajnyp{D. J. Gross}{Phys. Rev. Lett.}{32}{1974}{1071}.}
\item{6}{\ajnyp{E. G. Floratos, D. A. Ross and C. T.
 Sachrajda}{Nucl. Phys.}{B129}{1978}{66}; 
(E) {\bf B139} (1978) 545; 
\ajnyp{W. A. Bardeen et al.}{Phys. Rev.}{D18}{1978}{3998}.}
\item{7}{\ajnyp{W. L. van Neerven and E. B. Zijlstra}
{Phys. Lett.}{B272}{1991}{127 and 476}; 
{\sl ibid} { B273} {(1991)} {476}; {\sl Nucl. Phys.} {\bf B383} (1992) 525.}
\item{8}{\ajnyp{S. A. Larin et al.}{Nucl. Phys.}{B427}{1994}{41} and 
{\bf B492} (1997), 338.}
\item{9}{\kern0.25em{\smallsc A. L. Kataev, G. Parente and A. V. Sidorov},
IC/99/51 (hep-ph/9905310).}
\item{10}{\ajnyp{F. J. Yndur\'ain}{Phys. Lett.}{74B}{1978}{68}; see also  
\ajnyp{J. Wosiek and K. Zalewski}{Acta. Phys. Pol.}{B11}{1980}{697}.}
\item{11}{\ajnyp{L. W. Whitlow et al.}{Phys. Lett.}{B282}{1992}{475}; 
\ajnyp{A. Benvenuti et al.}{Phys. Lett.}{B223}{1989}{485};
\ajnyp{M. R. Adams et al.}{Phys. Rev.}{D54}{1996}{3006};
\ajnyp{M. Derrick et al.}{Z. Phys.}{C72}{1996}{399} and 
\ajnyp{S. Aid et al.}{Nucl. Phys.}{B470}{1996}{3}.}
\item{12}{\ajnyp{K. G. Chetyrkin et al.}{Phys. Rev. Lett.}{74}{1995}{2184}.}
\item{13}{\ajnyp{C. L\'opez and F. J. Yndur\'ain}{Nucl. Phys.}{B181}{1980}{231}; 
 {\rm ibid.,} {\bf B187} (1981) {157};\hb
 \ajnyp{K.~Adel, F. Barreiro and F. J. Yndur\'ain}
{Nucl. Phys.}{B495}{1997}{221}.}
\item{14}{\ajnyp{A.  Martin et al.}{Eur. Phys. J.}{C4}{1998}{463}.}
\item{15}{\ajnyp{S. Bethke}{Nucl. Phys. Suppl.}{B64}{1998}{54}.}
\item{16}{\ajnyp{G. Farrar}{Nucl. Phys.  Suppl.}{B62}{1998}{485}.}
\item{}{}
\item{}{}
\item{}{}
}
}\enddc

\bye